\documentclass[useAMS,usenatbib]{mn2e}
\usepackage{times}
\usepackage{dcolumn}
\usepackage{graphicx}
\usepackage{aas_macros}

\newcommand{\msun}{\mbox{${\rm M}_{\odot}$ }}
\newcolumntype{d}{D{.}{.}{-1}}

\title[Carbon-oxygen accretion discs in 4U~0614+09, 4U~1543-624 and
2S~0918-549]{Optical spectra of the carbon-oxygen accretion discs in
  the ultra-compact X-ray binaries 4U~0614+09, 4U~1543-624 and
  2S~0918-549}

\author[Nelemans, Jonker, Marsh \& van der Klis] {G.
  Nelemans$^{1}$\thanks{E-mail: nelemans@ast.cam.ac.uk, based on
    observations made with ESO Telescopes at the Paranal Observatories
    under programme ID 071.D-0119}, P.G. Jonker$^{1}$, T.R. Marsh$^{2}$ and M. van der Klis$^{3}$
  \\
  $^{1}$Institute of Astronomy, University of Cambridge, Madingley Road, Cambridge CB3 0HA, UK\\
  $^{2}$Department of Physics, University of Warwick, Coventry CV4
  7AL, UK\\
  $^{3}$Astronomical Institute ``Anton Pannekoek'', Univeristy of
  Amsterdam,
  Kruislaan 403, NL-1098 SJ Amsterdam, the Netherlands\\
}

\begin{document}

\date{Accepted . Received \today}

\pagerange{\pageref{firstpage}--\pageref{lastpage}} \pubyear{2003}

\maketitle

\label{firstpage}

\begin{abstract}
  We present optical spectra in the range 4600 -- 8600 \AA\ for three
  low-mass X-ray binaries which have been suggested to belong to the
  class of ultra-compact X-ray binaries based on their X-ray spectra.
  Our spectra show no evidence for hydrogen or helium emission lines,
  as are seen in classical X-ray binaries. The spectrum of 4U~0614+09
  does show emission lines which we identify with carbon and oxygen
  lines of C\textsc{ ii}, C\textsc{ iii}, O\textsc{ ii} and O\textsc{
    iii}. While the spectra of 4U~1543-624 and 2S~0918-549 have a
  lower signal-to-noise ratio, and thus are more difficult to
  interpret, some of the characteristic features of 4U~0614+09 are
  present in these spectra too, although sometimes they are clearly
  weaker. We conclude that the optical spectra give further evidence
  for the ultra-compact nature of these X-ray binaries and for their
  donor stars being carbon-oxygen white dwarfs.
\end{abstract}


\section{Introduction}\label{introduction}

Low-mass X-ray binaries are systems in which a neutron star or black
hole accretes from a low-mass companion. Most systems have orbital
periods of hours to days and are consistent with the scenario
\citep{heu83} in which the donors are main sequence or evolved,
hydrogen-rich, stars. A few ultra-compact systems have orbital periods
below an hour and are so compact that the donor stars cannot be main
sequence stars, but instead must be hydrogen poor
\citep[e.g.][]{vh95}.
  
From X-ray spectra of five X-ray binaries, including the persistent
sources 4U~0614+09, 4U~1543-624 and 2S~0918-549 \citet{jpc00} inferred
an enhanced neon/oxygen ratio, which they interpreted as being local
to the systems. The similarities between these three systems and the
other two, which are known ultra-compact X-ray binaries, led
\citet{jpc00} to conclude that these systems all have ultra-short
orbital periods and to propose that their donor stars originally were
carbon-oxygen or oxygen-neon-magnesium white dwarfs. Recently,
\citet{jc03} reported further X-ray spectroscopy for 4U~1543-624 and
2S~0918-549, confirming their earlier findings. \citet*{ynh02} argued,
based on mass-transfer stability arguments, that the donor stars in
ultra-compact X-ray binaries that have formed from white dwarf --
neutron star binaries should be low-mass white dwarfs ($M_{\rm donor}
\la 0.45 \msun$).  Combining binary evolution constraints and white
dwarf interior studies, \citet{ynh02} concluded that these systems
could be brought into one unifying scheme in which the systems were
descendants of binaries consisting of a so called hybrid white dwarf
\citep[carbon-oxygen (CO) core with thick helium mantle,][]{it85} and
a neutron star. The systems came into contact by angular momentum loss
due to gravitational wave emission, and quickly evolved from periods
of a few minutes to typical periods of tens of minutes. After cooling
for several Gyr the interior of a CO white dwarf crystallises and, due
to differential gravitational settling, chemical fractionation will
occur \citep[e.g.][]{her+94}.  At periods above 10 minutes neon
enriched layers could be exposed \citep[see][]{ynh02}. At these
orbital periods the only alternative donor stars are degenerate helium
stars or hydrogen poor remnants of stars that started mass transfer at
the very end of the main sequence \citep[for the latter,
see][]{phr01}, which would consist mainly of helium.
  

To further test the possible ultra-compact nature of 4U~0614+09,
2S~0918-549 and 4U~1543-624, we obtained optical spectra for these
sources, because ultra-compact systems are expected to be hydrogen
deficient, and even more, neon-rich donors stars are also expected to
be helium deficient. Previous optical spectra of 4U~0614+09
\citep{dms+74,mcc+90} indeed show no signs of the classical accretion
disc hydrogen emission lines but have a relatively low S/N ratio.

\section{Observations and reduction}\label{observations}

Spectra were taken with the FORS2 spectrograph on UT4 of the 8m Very
Large Telescope on Paranal in Chile. For each object we took spectra
both with the 1400V and 600RI holographic grisms, with a 1" slit,
using 2x2 on-chip binning.  This setup resulted in coverage of 4620 --
5930 \AA\ with mean dispersion of 0.64 \AA/pix for the 1400V spectra
and 5290 -- 8620 \AA\ with mean dispersion of 1.63 \AA/pix for the
600RI spectra. A log of the observations is given in Table~\ref{log}.

\begin{table}
\caption[]{Log of the observations}
\label{log}
\begin{tabular}{llllll}
\hline
 date, UT & grism & exp.time & airmass & seeing \\  
          &       & (s) &         &        \\\hline
\textbf{4U~0614+09} & & & & \\  
23/04/2003, 23:41:13 & 1400V & 2341 & 1.66 & 0.9 \\
24/04/2004, 23:30:29 & 600RI & 2341 & 1.61 & $\sim$1.3 \\
\textbf{4U~1543-624} & & & & \\
 25/04/2003, 09:10:06 & 1400V & 2661 & 1.47 & $\sim$0.5 \\
 27/04/2003, 06:17:24 & 1400V & 2301 & 1.26 & 1.3 \\
 25/04/2003, 08:17:31 & 600RI & 2661 & 1.36 & $\sim$0.5 \\ 
\textbf{2S~0918-549} & & & & \\
 27/04.2003, 23:57:33 & 1400V & 2661 & 1.16 & 0.5 \\
 30/04/2003, 01:31:59 & 1400V & 2301 & 1.26 & 1.3 \\
 30/04/2003, 00:40:22 & 600RI & 2661 & 1.19 & 0.7 \\
\hline
\end{tabular}
\end{table}

Data reduction was done using standard IRAF\footnote{IRAF is
  distributed by the National Optical Astronomy Observatories} tasks.
The bias was removed using the overscan region of the CCD, after which
the images were flatfield corrected using the standard calibration
plan flatfields. Spectra were extracted using optimal extraction
\citep{hor86} with the \texttt{apall} task. Arc lamp spectra were
extracted from the same place on the CCD. The 1400V wavelength
calibration was obtained using the positions of 17 lines, giving a
root-mean-square scatter of 0.05 \AA\ in fitting a fourth-order
Lagrangian polynomial.  The 600RI wavelength calibration uses 40 lines
and gives a root-mean-square scatter of 0.15 \AA\ for a fourth-order
Lagrangian polynomial.

The spectra were flux calibrated, using the nearest two flux standard
stars that were available from the VLT archive. Since these were taken
many days from our observations the flux calibration only provides a
very rough absolute flux calibration, but it does give a reasonable
estimate of the continuum shape. The reduced spectra were subsequently
imported in the MOLLY package for further analysis. For 4U~0614+09,
where the S/N ratio is the highest, we removed the telluric absorption
features by dividing the flux calibrated spectrum by a template of the
absorption. The template was constructed from the spectrum of a bright
star that was also in the slit as follows. We fitted a third order
cubic spline to line-free regions of the continuum of this spectrum
and used the fit to normalise the continuum to unity. The value of all
pixels in the spectrum was set to 1, except for the wavelength rages
6865--7700 and 8085 -- 8265 \AA, where strong telluric features are
present.

For each object all spectra were combined and averaged to obtain one
final spectrum.

\section{Analysis and interpretation}

\begin{figure*}
\begin{center}
\resizebox{2\columnwidth}{!}{\includegraphics[angle=-90,clip]{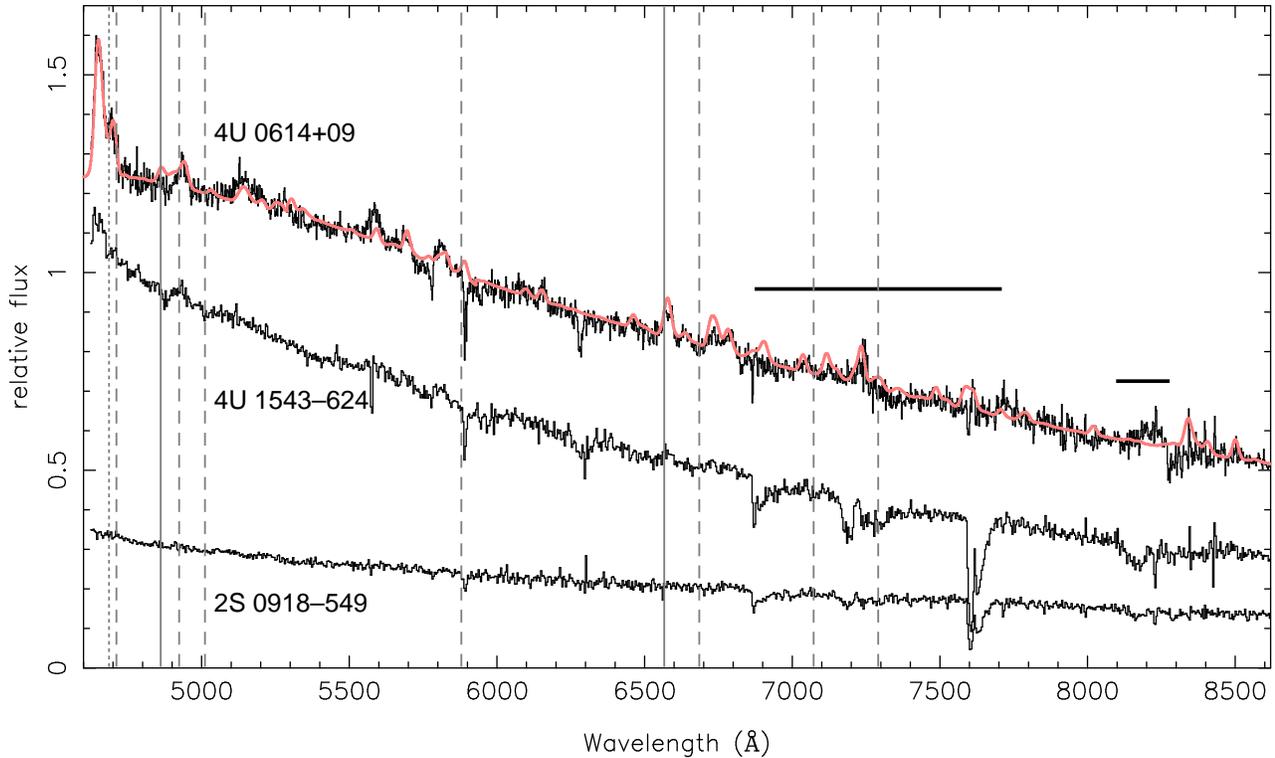}}
\end{center}
\caption[]{Spectra of 4U~0614+09, 4U~1543-624 and 2S~0918-549, rebinned to a scale of 2 or 4 \AA/pix. The thick smooth
  curve plotted over the 4U~0614+09 spectrum shows the pure carbon
  plus oxygen LTE model used to identify the lines, on top of fit to
  the continuum.  The vertical lines show the wavelength positions of
  several hydrogen (solid) and helium (dashed, dotted) emission lines
  seen in hydrogen or helium rich accreting systems. The horizontal
  bars show the wavelength ranges in which the 4U~0614+09 spectrum is
  corrected for telluric absorption.}
\label{fig:0614}
\end{figure*}

\subsection{Identification of the lines}\label{results}


In Fig.~\ref{fig:0614} we show the combined spectra. In the figure we
also show the positions of the most common hydrogen emission lines
seen in classical X-ray binaries as solid vertical lines (H$\alpha$ at
6563 \AA\ and H$\beta$ at 4861 \AA). We also show the positions of
several helium emission lines observed in optical spectra of some
interacting binaries \citep[e.g. GP~Com, see][]{mhr91}.  The dashed
lines show He\textsc{ i} lines at 4713, 4921, 5015, 5876, 6678, 7065
and 7286 \AA.  The dotted line shows the position of the He\textsc{
  ii} 4686 \AA\ line often seen together with hydrogen emission. From
Fig.~\ref{fig:0614} it is clear that the objects do not show strong
lines of hydrogen or helium. Upper limits on the equivalent widths
(EWs) of these lines are shown in Table~\ref{tab:other} (where we give
the value of $-$EW to avoid minus signs). There is a hint of
absorption around H$\beta$ in 4U~1543-624 and possibly emission around
He\textsc{ i} 4713 in 2S~0918-54.

\begin{table}
\caption{Negative equivalent width ($-$EW) upper limits (3 $\sigma$) on H and He lines, spectral shape and reddening
  determinations}
\label{tab:other}
\begin{tabular}{lrrr}
\hline
 & 4U~0614+09 & 4U~1543-624 & 2S~0918-549 \\ \hline
\multicolumn{4}{c}{\bf H and He $-$EWs (\AA)} \\
H$\beta$ (4861)  &  $< 0.33$ & $-$0.53 $\pm$ 0.1? & $< 0.36$ \\
He\textsc{ i} 5876         &  $< 0.22$ & $< 0.33$        & $< 0.36$ \\
He\textsc{ i} 6678         &  $< 0.41$ & $< 0.58$        & $< 1.0$  \\
He\textsc{ i} 7065         &  $< 0.40$ & $< 0.49$        & $< 0.73$ \\
He\textsc{ i} 5015         &  $< 0.26$ & $< 0.29$        & $< 0.35$ \\
He\textsc{ i} 4713         &  $< 0.24$ & $< 0.21$        & 0.41 $\pm$ 0.09? \\
He\textsc{ ii} 5411        &  $< 0.30$ & $< 0.46$        & $<0.61$ \\ 
\multicolumn{4}{c}{\bf Spectral shape} \\
$\alpha$ ($F_\lambda \propto \lambda^{-\alpha}$) & 1.67 (1.43) & 2.15 &
1.46 \\
\multicolumn{4}{c}{\bf Reddening measurements} \\
EW Na D lines & 1.67 $\pm$ 0.06 & 2.17 $\pm$ 0.08 & 1.90 $\pm$
0.1 \\
$\Rightarrow$ E(B$-$V) & $\ga 0.4$  & $\ga 0.4$ & $\ga 0.4$ \\
EW 5780 DIB  & 0.44 $\pm$  0.03 & 0.16 $\pm$ 0.05 & 0.35 $\pm$ 0.07
\\
$\Rightarrow$ E(B$-$V) & 0.88 $\pm$ 0.1 & 0.32 $\pm$ 0.1 & 0.70
$\pm$ 0.15 \\
E(B$-$V)$_{\rm max, SFK}$ & 0.51 & 0.32 & 0.6 \\
E(B$-$V)$_{\rm max, X}^a$  & 0.64 & 0.57 & 0.6 \\ \hline
\end{tabular}\\
$a$ From \citet{jpc00} using $N_H = 0.179 A_V 10^{22} {\rm cm}^{-2}$
\citep{ps95} and $A_V = 3.3 E(B-V)$ \citep{sfd98}.
\end{table}

The 4U~0614+09 spectrum shows clear, though sometimes weak, emission
lines.  We propose that these are due to partially ionised carbon and
oxygen, which have many lines in the optical region. In order to
identify the lines we calculated a simple emission profile for a 30-70
mixture (by number) of carbon and oxygen at a temperature of 27,000 K,
using an LTE emission line model, as described in \citet{mhr91}. We
assume a particle density of 3 $\times$ 10$^{13}$ cm$^{-3}$ and a line
of sight through the medium of 10$^{7}$ cm.  The only update is that
we consider ions up to C\textsc{ v} and O\textsc{ v} and use an atomic
line list compiled from the National Institute of Standards and
Technology\footnote{Version 2.0,
  http://physics.nist.gov/cgi-bin/AtData/lines\_form} and the Atomic
Line List\footnote{http://www.pa.uky.edu/$\sim$peter/atomic/}. The
model is smoothed with a 12 \AA\ Gaussian kernel, which at these
wavelengths corresponds to a $\sim$600 km s$^{-1}$ velocity width,
\emph{if} it is associated with kinematics. This model is not a
physical model of the accretion disc, as photoionisation is surely
important. However, we believe it correctly identifies the lines in
the observed spectrum.  Almost all features are blends of multiple
lines, most of them from C\textsc{ ii}, C\textsc{ iii} and O\textsc{
  ii}, while the strong feature at 5585 is due to O\textsc{ iii}.
Around 5810 C\textsc{ iv} might be present. In Table~\ref{tab:EWs} we
give the positions of the strongest features and the lines that may
contribute to them. We also list the measured equivalent widths of the
different features (again negative). The entries within parentheses
are either weak lines or uncertain identifications. The 7240 and 8220
features could be affected by the telluric correction.

The lower two spectra in Fig.~\ref{fig:0614} are of 4U~1543-624 and
2S~0918-549, rebinned to 4 \AA\ to enhance the S/N ratio. In
Table~\ref{tab:EWs} we also list the equivalent widths of the features
in these spectra.  The spectrum of 4U~1543-624 looks like a scaled
down version of the 4U~0614+09 spectrum, except maybe for the strength
of the C\textsc{ iii} and O\textsc{ ii} complexes around 4650 and 4700
\AA. All features, except the carbon features at 5280, 6070 and 6150
\AA\ are detected.  For 2S~0918-549 the case is less clear, mainly
because the lower signal. The features around 4650, 4700 and 6730 \AA\ 
are present, but most of the other features are undetected.

\begin{table*}
\caption{
Strongest features in the 4U~0614+09 spectrum and the (possible) line
identifications and measured negative EWs ($-$EWs, to avoid minus
signs) for the three spectra. For each feature the wavelength range
over which the EW is determined is given in parenthesis.
}
\label{tab:EWs}
\begin{minipage}{\columnwidth}
\hspace*{-0.2cm}
\begin{tabular}{@{}l@{\hspace{0.15cm}}l@{\hspace{0.15cm}}l@{\hspace{0.35cm}}l@{\hspace{0.35cm}}l@{\hspace{0.35cm}}l@{}} \hline
Feature  & Ion     &    lines   &        & $-$EW (\AA)             &   \\
 & & &  \hspace*{-0.1cm}\textbf{4U~0614+09} & \hspace*{-0.1cm}\textbf{4U~1543-624} & \hspace*{-0.1cm}\textbf{2S~0918-549} \\\hline 
\textbf{4650}     & C\textsc{ iii}    &    4647.418        & 9.77 $\pm$ 0.15 & 3.34 $\pm$ 0.13 &  1.99 $\pm$ 0.15 \\
(4624-    &         &    4650.246        &               &\\
4680)     &         &    4651.016        &                 &\\
         &         &    4651.473        &               &\\
         &         &    4652.048        &               &\\
         &         &    4659.058        &               &\\
         &         &    4663.642        &               &\\
         &         &    4665.860        &               &\\
         &         &    4670.49         &               &\\
         &         &    4671.22         &               &\\
         &         &    4671.74         &               &\\
         &         &    4673.953        &               &\\
         & O\textsc{ ii}     &    4638.8558       &               &\\
         &         &    4641.8103       &               &\\
         &         &    4649.1347       &               &\\
         &         &    4650.8384       &               &\\
         &         &    4661.6324       &               &\\
         &         &    4673.7331       &               &\\
         &         &    4676.2350       &               &\\ \hline
\textbf{4700}     & O\textsc{ ii}     &    4690.888        &3.51 $\pm$  0.12& 0.80 $\pm$ 0.11& 1.37 $\pm$ 0.13 \\
(4680-    &         &    4691.419        &         &\\
4720)     &         &    4698.437        &          &\\
         &         &    4699.011        &               &\\
         &         &    4699.218        &               &\\
         &         &    4701.179        &               &\\
         &         &    4701.712        &               &\\
         &         &    4703.161        &               &\\
         &         &    4705.346        &               &\\
         &         &    4710.009        &               &\\ \hline
\textbf{4935}     & O\textsc{ ii}     &    4906.830        & 1.79 $\pm$ 0.11   & 1.04 $\pm$ 0.12  &    0.54 $\pm$ 0.13       \\
(4900-    &         &    4924.529        &              &\\
4960)     &         &    4941.072        &                &\\
         &         &    4943.005        &               &\\
         &         &    4955.707        &               &\\ \hline
\textbf{5140}     & C\textsc{ ii}     &    5132.947        & 2.40 $\pm$ 0.097&  1.12 $\pm$ 0.10&  0.068 $\pm$ 0.13       \\
(5110-    &         &    5133.282        &&\\
5165)     &         &    5143.495        &              &\\
         &         &    5145.165        &               &\\
         &         &    5151.085        &               &\\
         & (O\textsc{ ii})   &    5159.941        &               &\\ \hline
\textbf{5190}     & O\textsc{ ii}     &    5175.903        & 1.54 $\pm$ 0.099 &   0.80 $\pm$ 0.11  &  0.38 $\pm$ 0.14       \\
(5160-    &         &    5190.498        &       &               \\
5220)     &         &    5206.651        &         &               \\ \hline
\textbf{5280}     & C\textsc{ iii}    &    5249.112        & 0.98 $\pm$ 0.11       &    0.36 $\pm$ 0.14 &   $-$0.15 $\pm$ 0.19      \\
(5230-    &         &    5253.575        &         &               \\
5310)     &         &    5272.522        &         &               \\
         & (O\textsc{ iii})  &    5268.301        &       &               \\
         & (C\textsc{ ii})   &    5257.236        &       &               \\
         &         &    5259.056        &       &               \\
         &         &    5259.664        &       &               \\
         &         &    5259.758        &       &               \\ \hline
\end{tabular}
\end{minipage}
\begin{minipage}{\columnwidth}
\hspace*{0.15cm}
\begin{tabular}{@{}l@{\hspace{0.15cm}}l@{\hspace{0.15cm}}l@{\hspace{0.35cm}}l@{\hspace{0.35cm}}l@{\hspace{0.35cm}}l@{}} \hline
Feature  & Ion     &    lines   &        & $-$EW (\AA)             &   \\
 & & &  \hspace*{-0.1cm}\textbf{4U~0614+09} & \hspace*{-0.1cm}\textbf{4U~1543-624} & \hspace*{-0.1cm}\textbf{2S~0918-549} \\\hline 
\textbf{5585}     & O\textsc{ iii}    &    5592.252        & 3.12 $\pm$ 0.12       &    2.11 $\pm$ 0.18  &  1.38 $\pm$ 0.23       \\
(5550-    & (OV??)  &    (5580.12)       &         &               \\
5620)     &         &    (5597.89)       &          &               \\ \hline
\textbf{5700}     & C\textsc{ iii}    &    5695.92         &  1.21 $\pm$ 0.095     &    1.18 $\pm$ 0.15    &  0.44 $\pm$ 0.19         \\
(5675-    &         &            &       &            \\
5715)     &         &            &       &                  \\ \hline
\textbf{5810}     & C\textsc{ iii}    &    5826.42         & 1.76 $\pm$ 0.11       & 1.29 $\pm$ 0.17 &  0.71 $\pm$ 0.25              \\
(5785-    & C\textsc{ iv}?    &    5801.31         &        &               \\
5840)     &         &    5811.97         &    &               \\ \hline
\textbf{6070}     & C\textsc{ ii}?    &    6095.29         & 0.61 $\pm$ 0.23       &   0.77 $\pm$ 0.33 &  $-$1.52 $\pm$ 0.52            \\
(6040-    &         &    6098.51         &        &               \\
6120)     &         &            &       &                 \\ \hline
\textbf{6150}     & C\textsc{ ii}     &    6151.27         & 0.56 $\pm$ 0.19       &   0.21 $\pm$ 0.25   &  0.79 $\pm$ 0.42           \\
(6130-    &         &    6151.54         &     &               \\
6180)     & (C\textsc{ iii})  &   (6155.12)        &         &               \\
         &         &   (6156.69)                &       &               \\ \hline
\textbf{6580}     & C\textsc{ ii}     &    6578.05         & 3.14 $\pm$ 0.15       &  1.72 $\pm$ 0.20 &  0.55 $\pm$ 0.37     \\
(6550-    &         &    6582.88         &        &               \\
6600)     & (O\textsc{ ii})   &    (6565.283)      &        &               \\
         &         &    (6571.108)      &       &               \\ \hline
\textbf{6730}     & C\textsc{ iii}    &    6727.48         & 2.18 $\pm$ 0.16       &  1.91 $\pm$ 0.21 &  1.33 $\pm$ 0.38          \\
(6700-    &         &    6731.04         &        &               \\
6760)     &         &    6742.15         &        &               \\
         &         &    6744.39         &       &               \\
         & O\textsc{ ii}     &    6717.754        &       &               \\
         &         &    6721.388        &       &               \\
         & (C\textsc{ ii}?)  &    many weak       &       &               \\ \hline
\textbf{6790}     & C\textsc{ ii}     &    6779.94         & 2.52 $\pm$ 0.15       &  2.42 $\pm$ 0.21 &  3.83 $\pm$ 0.38             \\
(6760-    &         &    6780.59         &         &               \\
6820)     &         &    6783.91         &         &               \\
         &         &    6787.21         &       &               \\
         &         &    6791.47         &       &               \\
         &         &    6798.10         &       &               \\
         &         &    6800.69         &       &               \\
         & C\textsc{ iii}    &    6774.95         &       &               \\ \hline
\textbf{(7240)}   & (C\textsc{ ii})   &    (7231.33)       & (2.97 $\pm$ 0.20)     &          \\
(7210-    &         &    (7236.42)       &       &               \\
7260)     &         &    (7237.17)       &       &               \\ \hline
\textbf{7720?}    & C\textsc{ iii}?   &    7707.43         & 1.75 $\pm$ 0.23       &   0.62 $\pm$ 0.28 &  $-$1.27 $\pm$ 0.38            \\
(7700-    &         &            &       &                \\
7750)     &         &            &       &               \\ \hline
\textbf{(8220)}   & (O\textsc{ i}?)   & (8230)             & (4.1 $\pm$ 0.34)      &              \\
(8180-    &         &            &       &               \\
8260)     &         &            &       &               \\
         &         &            &       &               \\ \hline
\vspace*{1.67cm}
\end{tabular}
\end{minipage}
\end{table*}

\subsection{Spectral shape and reddening}

In Table~\ref{tab:other} we also list the measured spectral slope of
the three objects, assuming a power law ($F_\lambda \propto
\lambda^{-\alpha}$), and the temperatures of the best fitting black body
spectra ($T_{\rm BB}$). The spectrum of 4U~1543-624 is significant
bluer than the other two. For 4U~0614+09, the continuum is well fit
with $\alpha = 1.67$ for $\lambda \ga 5100$ \AA, but is significant
flatter below. This flattening is not seen in the spectrum of
\citet{mcc+90}, and thus might be due to the flux calibration.

We also list several estimates or limits on the interstellar reddening
of the objects. We have two possible reddening indicators available in
the spectra: the Na D lines and the diffuse interstellar band (DIB) at
5780 \AA. In Table~\ref{tab:other} we list the measure equivalent
widths of these features and the implied reddening according to
\citet{mz97} and \citet{her93} respectively.  The Na D lines only give
rough lower limits, as the EWs level off above $E(B-V) \approx$ 0.5
due to saturation. The high EWs measured here would then be the result
of multiple components in the lines, as can indeed be seen in the
4U~0614+09 spectrum. In the Table we also list the maximum reddening
according to the \citet{sfd98} dust maps and the implied reddening for
the hydrogen column found in the spectral modelling of the X-ray
spectra \citep{jpc00}. These last values are upper limits, as the
local O and Ne absorption implied by the spectra would artificially
enhance the hydrogen column in the fits to the X-ray spectra
\citep[e.g.][]{jpc00}. Although the above gives a barely consistent
picture, we conclude that the reddening of these objects is probably
close to the maximum in the Galaxy in their directions.

\vspace*{-0.3cm}
\subsection{Origin of the line emission}

Although detailed modelling of the observed spectra is beyond the
scope of this paper, we briefly discuss the possible origin of the
line emission. The spectral shapes of the continua, corrected for the
reddening (using $E(B-V)$ = 0.6, 0.4 and 0.6 for 4U~0614+09,
4U~1543-624 and 2S~0918-549 respectively) are consistent with a black
body of temperature $\sim$20,000 K. The fact that these objects are
persistent X-ray sources, could point at origin of the lines in either
the irradiated disc or the irradiated donor star, although in that
case more than doubly ionised species might be expected to be present.
Only phase resolved spectroscopy will conclusively tell, but the
observed width of the lines ($\sim600$ km s$^{-1}$) might suggest the
inner parts of the accretion disc.

\section{Discussion}\label{discussion}

Previous spectra have been interpreted in a different way to that
presented here, but were hampered by quite low S/N ratio.  For
instance \citet{mcc+90}, interpreted the strongest lines near 4650
\AA\ and 5590 \AA, as the Bowen blend and possibly O\textsc{ i} at
5577 \AA.  They already commented on the intriguing absence of the
usual He~II 4686 \AA\ line, which casts doubt on the interpretation of
the emission at 4650 \AA\ as the Bowen blend, since the Bowen
mechanism is driven by helium. The total absence of any sign of helium
at any of the positions indicated in the plots strongly argues against
(much) helium in the system.

As discussed in the Introduction, \citet{jpc00} suggested the donor
stars originally were carbon-oxygen or oxygen-neon-magnesium white
dwarf, while \citet{ynh02} argued they were hybrid white dwarfs.  Our
spectra thus give further evidence for the interpretation of
4U~0614+09, 2S~0918-549 and 4U~1543-624 as ultra-compact X-ray
binaries with carbon-oxygen white dwarf donors. One very interesting
prospect is the determination of the carbon to oxygen ratio in the
transferred material from detailed modelling of the disc spectrum,
which would give an unprecedented view into the interior of a white
dwarf and possibly could be used to constrain the rate of the
$^{12}C(\alpha, \gamma)^{16}O$ reaction \citep[see for a discussion
and references][]{sdi+03}.

Further evidence for the ultra-compact nature of 4U~0614+09 and
2S~0918-549 comes from their absolute visual magnitude. Using the
magnitudes and reddening of the objects from the low-mass X-ray binary
catalogue \citep{lvv01} and the distances of these systems of $<$3 and
4.2 kpc respectively from type I X-ray bursts \citep{bcl+92,cvi+02} we
find absolute magnitudes of $>$5.4 and 6.9 for 4U~0614+09 and
2S~0918-549. These are very faint for X-ray binaries and, from the
\citet{vm94} relation between absolute magnitude, X-ray luminosity and
orbital period, suggest periods well below one hour. The derived X-ray
luminosities of $\la 0.01 L_{\rm Edd}$ for 4U~0614+09 \citep{fkt+96}
and $\sim0.003 L_{\rm Edd}$ for 2S~0918-549 \citep{jvh+01},
assuming these are indicative of the average mass accretion rate, give
mass transfer rates of $\sim$3 $\times$ 10$^{-10}$ and $\sim$9
$\times$ 10$^{-11}$ \msun yr$^{-1}$ for 4U~0614+09 and 2S~0918-549,
assuming accretion of helium in the type I X-ray bursts (see below).
Comparing these rates with Fig. 15 of \citet{db03}, suggests orbital
periods of 15-20 and 20-30 minutes for 4U~0614+09 and 2S~0918-549.

Finally we note that our findings pose an interesting question
concerning type I bursts. As discussed by \citet{jc03}, the bursts
observed in 4U~0614+09 \citep{sbh+78,bcl+92} and 2S~0918-549
\citep{jvh+01} are all short bursts, which are believed to be caused
by hydrogen and/or helium. They suggest that possibly the donor stars
still have non-negligible hydrogen fraction, which would have evolved
from binaries that start mass transfer close to the end of the main
sequence \citep{nrj86,prp02}. However, these are expected to consist
mainly of helium, rather than carbon and oxygen. The alternative
\citet{jc03} suggest, which seems the only remaining option in light
of the lack of helium lines in the optical spectrum, is that
spallation of the carbon and oxygen nuclei at the impact onto the
neutron star \citep[cf.][]{bsw92} turns them into helium (or possibly
hydrogen), which subsequently triggers the burst.

\section{Conclusions}\label{conclusions}

We presented optical spectra of the three suspected ultra-compact
X-ray binaries 4U~0614+09, 4U~1543-624 and 2S~0918-549. The spectra
show no sign of hydrogen or helium emission lines. We identify the
observed features as lines of C\textsc{ ii}, C\textsc{ iii},
O\textsc{ ii} and O\textsc{ iii}. This is in agreement with the
interpretation of these sources as ultra-compact X-ray binaries and
the expectation that the donor stars in these objects originally were
hybrid white dwarfs that have lost most of their mass and now consists
mainly of oxygen and carbon. These spectra are thus of accretion discs
almost purely made out of oxygen and carbon.

\section*{Acknowledgments}

We thank the referee Janet Drew for comments that improved the paper.
We are thankful to Peter van Hoof and the National Institute of
Standards and Technology for compiling the atomic line lists we use.
We further thank Lars Bildsten for stimulating discussions. GN
acknowledge the hospitality of the Kavli Institute for Theoretical
Physics. This work was supported by the National Science Foundation
under grant PHY99-07949.

\bibliography{journals,binaries} \bibliographystyle{mn2e}

\label{lastpage}
\end{document}